\documentclass[conference]{IEEEtran}

\usepackage{cite}
\usepackage{amsmath,amssymb}
\usepackage{graphicx}
\usepackage{booktabs}
\usepackage{multirow}
\usepackage{threeparttable}
\usepackage{array}
\newcolumntype{M}{>{\centering\arraybackslash}m{0.73cm}}
\usepackage{url}
\usepackage{needspace}
\usepackage{eso-pic}

\begin{document}

\AddToShipoutPictureFG*{%
  \AtPageLowerLeft{%
    \raisebox{0.14in}[0pt][0pt]{%
      \makebox[\paperwidth][c]{%
        \parbox{0.92\paperwidth}{\centering\scriptsize
        \textcopyright~2026 IEEE. Personal use of this material is permitted. Permission from IEEE must be obtained for all other uses, in any current or future media, including reprinting/republishing this material for advertising or promotional purposes, creating new collective works, for resale or redistribution to servers or lists, or reuse of any copyrighted component of this work in other works.}%
      }%
    }%
  }%
}

\title{HSMLog: Small Language Model-Assisted \\ Hardware Security Module Log Anomaly Detection with Behavioral Analysis}

% ==============================================================================================================

\author{
\IEEEauthorblockN{
Chia-Hsuan Wu\IEEEauthorrefmark{1},
Dar-Hsin Dustin Wu\IEEEauthorrefmark{1},
Rui Fang\IEEEauthorrefmark{1},
Yi-Ting Lee\IEEEauthorrefmark{1},
Chia-Chih Lin\IEEEauthorrefmark{2},
and Ming-Syan Chen\IEEEauthorrefmark{1}
}
\IEEEauthorblockA{
\IEEEauthorrefmark{1}National Taiwan University, Taipei, Taiwan\\
\IEEEauthorrefmark{2}National Taiwan University of Science and Technology,
Taipei, Taiwan\\
Email: \{chwu, dustinwu, rfang, ytlee\}@arbor.ee.ntu.edu.tw,
cclin@mail.ntust.edu.tw, mschen@ntu.edu.tw
}
}

\maketitle

% ====================================================================================

\begin{abstract}
Hardware Security Module (HSM) logs capture security-critical behavior, but
anomalies emerge from relationships across event sequences, keys, object
states, sessions, and temporal patterns rather than isolated events. Existing
methods separate detection from HSM-specific evidence validation and reporting.
In this paper, we present HSMLog, a two-stage framework for HSM log anomaly
detection with retrieval-grounded behavioral analysis. In Stage 1, a small
language model (SLM) identifies candidate alerts from sliding windows of
structured HSM events and performs policy-guided assessment using HSM-specific
operational rules. In Stage 2, retrieved policies and historical suspicious-key
records strictly predating the alert window, together with candidate-related
log context, support conservative candidate review and incident analysis.
Evaluated on real industrial HSM background logs augmented with anomaly
scenarios co-defined with industrial partners, HSMLog achieves 98.97\% precision,
96.00\% recall, 98.66\% anomalous-event coverage, and a 97.46\% F1 score,
demonstrating effective anomaly alerting and incident triage in the studied
setting.
\end{abstract}

% ==============================================================================================================

\section{Introduction}

Hardware Security Modules (HSMs) protect cryptographic keys and execute
security-critical operations, including signing, encryption, decryption, key
generation, object management, and authentication. Failures, misuse, and
behavior that departs from deployment expectations can affect security, service
availability, auditability, and the reliability of dependent applications.
Detecting such conditions is therefore important for maintaining dependable
HSM-backed services.

Operational HSM logs differ from conventional system logs because each record
describes a stateful cryptographic operation rather than only an event template
or message. Its meaning depends on the target key or object, the session and
role that issued the request, the object's lifecycle state, the cryptographic
mechanism, returned handles, and the result code. For example, an
invalid-handle result may be expected after a stale reference but may indicate
object exploration when repeated across unrelated objects. Likewise, a
successful operation may still require investigation when it violates a
deployment-specific policy. Consequently, behavior that appears benign at the
event level may become suspicious only when interpreted as part of a structured
operational progression.

Existing log anomaly detection methods have achieved promising results on
public datasets~\cite{Du2017DeepLog,Guo2021LogBERT,Zhang2019LogRobust}, but
their application to operational HSM environments remains challenging.
Sequence-only detectors can overlook organization-specific policies, whereas
policy-guided SLM assessment alone cannot capture every contextual workflow
deviation. Either alone may miss repeated operations involving the same
protected asset, requester, or invalid state transition, all central to HSM
interpretation.
Industrial settings rarely provide exhaustive event-level
labels, with supervision often limited to annotated cases or incident spans.
Given sensitive HSM logs, HSMLog uses a compact SLM over local windows, avoiding
full-history inference; Stage~2 retrieves broader evidence only for detected candidates.

This paper presents HSMLog, a two-stage framework for HSM log anomaly detection 
with retrieval-grounded behavioral analysis. Stage~1 uses an SLM to analyze structured 
event windows and assess organization-specific policies. Positive windows from contextual 
detection or policy-guided assessment are merged into candidate incidents. Stage~2 retrieves 
policies, suspicious-key records strictly predating the incident interval, and candidate-related 
log context, supporting conservative review and evidence-grounded behavioral analysis.

The contributions of this paper are as follows:

\begin{itemize}
\item We propose HSMLog, a two-stage framework that combines SLM-based
contextual anomaly detection and policy-guided SLM assessment of
organization-specific operational policies with retrieval-grounded candidate
review and evidence-grounded behavioral analysis.

\item We develop a structured representation of HSM events and a case-aware
sliding-window training protocol that enable contextual SLM detection from
annotated operational scenarios.

\item We evaluate HSMLog on real operational logs from an industrial HSM
environment using five practitioner-informed anomaly categories:
Cryptographic Failure Probing, Object Handle Probing, Lifecycle Violation,
Policy Violation, and Low-and-Slow Metadata Probing. We compare HSMLog with
representative baselines and analyze case detection, anomalous-event coverage,
component contributions, retrieval-supported candidate review, and
observation-window sensitivity.
\end{itemize}

% ==============================================================================================================

\section{Industrial Context and Problem}

\subsection{Hardware Security Modules and the PKCS\#11 Interface}

In the studied environment, applications access HSM services through PKCS\#11
(Cryptoki), a standard cryptographic token interface~\cite{PKCS11}. Unlike
ordinary system logs, HSM audit records expose structured security-relevant
entities: key or object handles and labels identify protected targets; session
identifiers and roles indicate requesting authority; attributes and lifecycle
operations describe object state; mechanisms specify cryptographic actions; and
returned handles and result codes record outcomes. These fields must be
interpreted jointly to determine whether an operation is consistent with an
expected cryptographic workflow.

\subsection{Operational HSM Monitoring Challenges}

An HSM operation may be expected in one workflow but suspicious when its target,
state, authority, or outcome conflicts with surrounding activity or policy.
Anomaly evidence is therefore sparse, relational, and state-dependent, and not
every contextual deviation can be specified in advance.

This study uses anonymized industrial HSM audit data and anomaly scenarios
co-defined with industrial partners. The partners helped identify relevant log
fields and triage-oriented behavioral conditions, with sensitive content removed
while preserving the event schema and behavioral structure.

% ==============================================================================================================

\subsection{Problem Definition}

Let $\mathcal{E}=(e_1,e_2,\ldots,e_N)$ denote a canonicalized HSM audit-event
stream. HSMLog constructs a local observation window
$W_j=(e_j,e_{j+1},\ldots,e_{j+w-1})$ and performs first-stage candidate
detection as

\begin{equation}
z_j =
d_{\mathrm{context}}(W_j)
\lor
d_{\mathrm{policy}}(W_j),
\qquad
z_j \in \{0,1\},
\end{equation}

where $d_{\mathrm{context}}$ denotes the SLM's contextual anomaly decision and
$d_{\mathrm{policy}}$ denotes its policy-guided assessment against applicable
organization-specific operational policies.

Related positive windows are merged into candidate incidents to avoid
duplicated alerts:

\begin{equation}
\mathcal{C} =
\operatorname{Merge}\!\left(
\left\{ W_j \mid z_j = 1 \right\}
\right).
\end{equation}

For each candidate $c_k \in \mathcal{C}$, Stage~2 retrieves operational policies $P_k$ and suspicious-key records $H_k^{-}$ that strictly predate the candidate incident interval, and collects candidate-related log context $L_k$.
 The retrieved evidence and associated log context support
conservative candidate review and, for retained candidates, retrieval-grounded
behavioral analysis:

\begin{equation}
\hat{a}_k =
v_{\mathrm{review}}(c_k, P_k, H_k^{-}, L_k),
\qquad
\hat{a}_k \in \{0,1\}.
\end{equation}

Here, $\hat{a}_k=1$ indicates that candidate $c_k$ is retained for
investigation; otherwise, it is not retained after evidence review. For each
retained candidate, HSMLog generates a retrieval-grounded behavioral analysis
$r_k$ from the retrieved evidence and associated log context. The objective is
to identify suspicious local behavior and retain candidates with sufficient
operational support for engineer investigation.

% ==============================================================================================================

\section{HSMLog Approach}

\subsection{Framework Overview}

\begin{figure*}[t]
\centering
\includegraphics[
  width=\textwidth,
  trim=28bp 8bp 16bp 8bp,
  clip
]{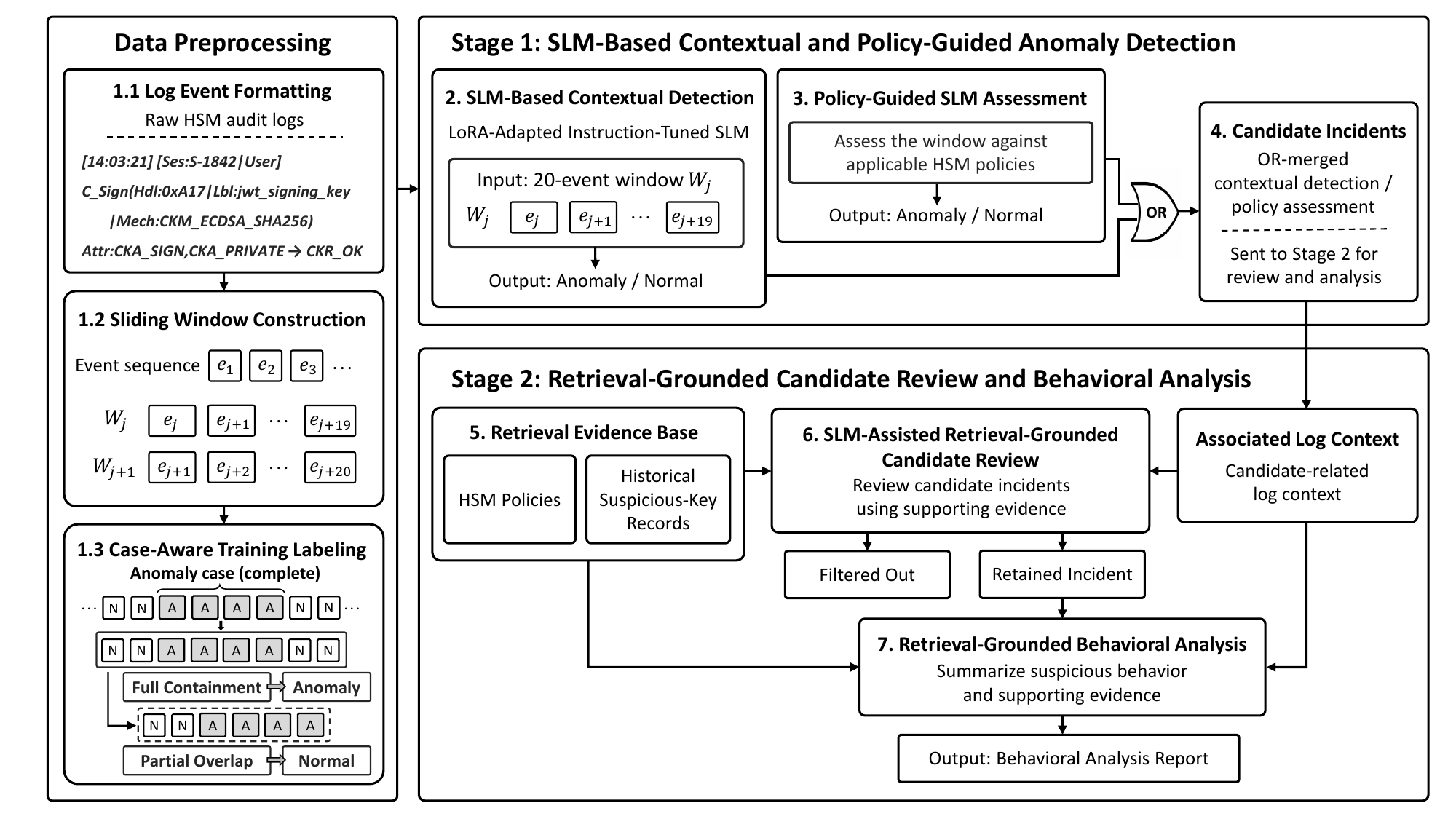}
\caption{Overview of HSMLog. Stage~1 performs SLM-based contextual and
policy-guided anomaly detection over structured HSM windows. Stage~2 uses retrieved policies, prior
suspicious-key records, and candidate-related log context for conservative
review and retrieval-grounded behavioral analysis.}
\label{fig:framework}
\end{figure*}

Figure~\ref{fig:framework} presents the two-stage HSMLog workflow.

After preprocessing constructs overlapping structured event windows, Stage~1
classifies each window as \textit{Anomaly} or \textit{Normal} and assesses it
against applicable organization-specific policies from protected HSM knowledge,
without disclosing confidential policy content or criteria. Positive windows
from contextual detection or policy-guided assessment are merged into candidate
incidents. This design combines learned workflow patterns with explicit
operational constraints, covering both workflow deviations and policy-level
misuse.

Stage~2 applies only after a candidate incident is identified. It retrieves HSM policies, suspicious-key records strictly predating the incident interval, and candidate-related events for conservative review and retrieval-grounded behavioral analysis. It introduces no new detections, and this temporal restriction prevents the incident from supporting itself.

% ==============================================================================================================

\subsection{Structured HSM Event Representation}

Raw HSM audit records contain interdependent fields whose independent treatment
can obscure relationships among operations, keys, objects, sessions, and
results. HSMLog converts each record into a compact, behavior-preserving
representation for SLM inference.

\begin{figure}[!t]
\centering
\includegraphics[
  width=\columnwidth,
  trim=192bp 1bp 191bp 1bp,
  clip
]{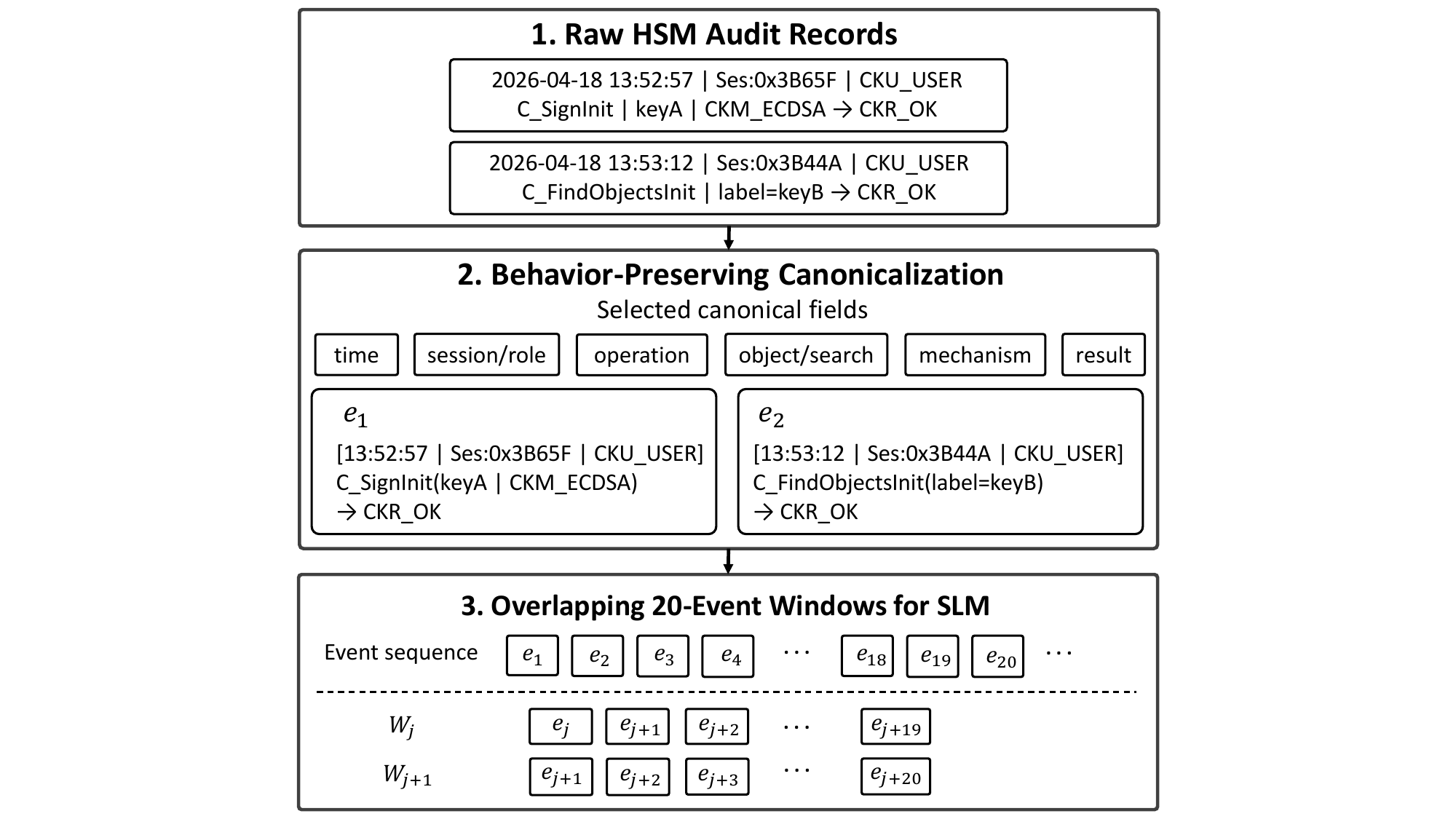}
\caption{Behavior-preserving preprocessing for HSM logs. Raw records are
canonicalized into compact events and organized into overlapping event windows
for SLM classification. Only selected log fields are shown for clarity.}
\label{fig:preprocessing}
\end{figure}

Figure~\ref{fig:preprocessing} illustrates this preprocessing pipeline.

The representation retains timestamps, session IDs, roles, operations, key
handles and labels, mechanisms, attributes, search templates, returned handles,
and result codes. List or dictionary attributes are normalized to text:

\par\bigskip
\Needspace{4\baselineskip}
{\footnotesize\ttfamily
\noindent
\begin{tabular}{@{}l@{}}
[time]\ [Ses:session|role] \\
operation(Hdl:handle|Lbl:label|Mech:mechanism) \\
{}[Attr:attributes|Srch:template] \\
\quad $\rightarrow$ [Ret:handles|Result:result]
\end{tabular}\par}
\bigskip

In this representation, \texttt{Hdl} and \texttt{Lbl} link operations on the
same key or object; \texttt{Ses} and role preserve requesting context; and
\texttt{Mech}, \texttt{Attr}, \texttt{Ret}, and \texttt{Result} preserve
operation intent and outcome. This keeps security-relevant fields connected
rather than treating them as independent tokens. The event stream enables SLM
reasoning over object-state progression, cross-object access, repeated
failures, and session-dependent usage beyond operation names.

% ==============================================================================================================

\subsection{Sliding Windows and Case-Aware Training Protocol}

HSMLog uses overlapping local event windows over the canonical event stream, so
related behavior can remain visible even when benign activity is interleaved,
without assuming a single application session.

For training, labels follow annotated scenario spans. A window is labeled
\textit{Anomaly} only when it fully contains an annotated case; partial overlaps
remain \textit{Normal}. This requires complete behavioral evidence but may place
anomalous events in Normal-labeled windows near case boundaries, reflecting a
conservative labeling trade-off.

% ==============================================================================================================

\subsection{Stage~1: SLM-Based Contextual Detection}

HSMLog uses an instruction-tuned small language model to classify each
canonicalized HSM event window as \textit{Anomaly} or \textit{Normal} using an
HSM-specific instruction. The instruction directs the
model to assess sequence coherence and contextual consistency across events,
sessions, keys, objects, and outcomes, including object-state progression,
access relationships, session-dependent behavior, mechanism usage, and
result-code patterns.

This enables HSMLog to identify behavior that appears benign in isolation but
becomes suspicious through repeated failures, cross-object access, lifecycle
inconsistencies, or other HSM-specific context.

% ==============================================================================================================

\subsection{Stage~1: Policy-Guided SLM Assessment}

Some behaviors depend on organization-specific policies beyond contextual
detection. For each window, HSMLog supplies the SLM with applicable protected
rules and log context for policy-guided assessment, then combines this decision
with contextual evidence. The same rules support Stage 2 review and
explanation, not new detection; confidential details remain undisclosed.
To avoid duplicate alerts, overlapping positive windows are merged into a
single candidate incident.

% ==============================================================================================================

\subsection{Stage~2: Retrieval-Grounded Candidate Analysis}

Stage~2 provides post-detection analysis using Retrieval-Augmented Generation
(RAG)~\cite{Lewis2020RAG}. For each candidate incident, HSMLog uses its
implicated keys, operations, and time interval to retrieve relevant policies
and strictly earlier suspicious-key records, while collecting candidate-related
events from the merged incident span. These inputs support conservative review:
Stage~2 filters only weak, unsupported candidates, introduces no new detections,
and uses no ground-truth labels.

For each retained incident, HSMLog organizes the same evidence into a
retrieval-grounded behavioral analysis that identifies affected keys or
objects, summarizes suspicious progression, states supporting evidence, and
highlights investigation-relevant context for post-detection explanation and
triage.

% ==============================================================================================================

\section{Evaluation Setup}

\subsection{Dataset Construction}

The evaluation combines real operational HSM logs with anomaly scenarios
co-defined with industrial partners. Training contains 27,319 events: 24,582 normal-background events from 352
sessions and 2,737 anomalous events from 180 labeled cases. The
contextual SLM is trained on four categories; Policy Violation is assessed only
through policy-guided assessment at inference.

The final test stream contains 16,573 events: 15,001 normal-background events
and 1,572 scenario-instantiated anomalous events across 100 near-balanced cases
in five categories. Normal backgrounds are session-disjoint, and training and
final-test anomaly instances are independently instantiated with disjoint
sessions, keys, handles, and event windows.

% ==============================================================================================================

Table~\ref{tab:data_composition} summarizes the anomaly composition.

\begin{table}[t]
\centering
\caption{Training and final-test anomaly composition.}
\label{tab:data_composition}
\footnotesize
\renewcommand{\arraystretch}{1.08}
\setlength{\tabcolsep}{1.2pt}

\begin{tabular*}{\columnwidth}{
@{\extracolsep{\fill}}
>{\raggedright\arraybackslash}p{0.53\columnwidth}
r r r r
@{}
}
\toprule
\multirow{2}{*}{\textbf{Category}} &
\multicolumn{2}{c}{\textbf{Train}} &
\multicolumn{2}{c}{\textbf{Test}} \\
\cmidrule(lr){2-3}
\cmidrule(lr){4-5}
& \textbf{Cases} & \textbf{Events}
& \textbf{Cases} & \textbf{Events} \\
\midrule
Cryptographic Failure Probing
& 45 & 732 & 21 & 344 \\

Object Handle Probing
& 45 & 736 & 19 & 313 \\

Lifecycle Violation
& 45 & 577 & 20 & 258 \\

Policy Violation
& 0 & 0 & 21 & 363 \\

Low-and-Slow Metadata Probing
& 45 & 692 & 19 & 294 \\
\midrule
\textbf{Total}
& \textbf{180} & \textbf{2,737}
& \textbf{100} & \textbf{1,572} \\
\bottomrule
\end{tabular*}
\end{table}

% ==============================================================================================================

\subsection{Implementation Details}
\label{sec:implementation}

Stage~1 uses a predefined observation window of $w=20$, fixed before final-test
evaluation to preserve broader context for operations separated or interleaved
by routine activity. The contextual detector uses
\texttt{google/gemma-4-E4B-it}~\cite{GoogleDeepMind2026Gemma4} with
LoRA~\cite{Hu2022LoRA} applied to linear modules with rank $r=16$, scaling
factor $32$, and dropout $0.05$.

Fine-tuning processes 78.85 million tokens across 54,600 window instances
(two training passes over 27,300 windows), using a learning rate of
$2\times10^{-5}$ and response-only supervision with loss applied only to
classification completions. This focuses optimization on the classification
decision rather than reproducing the input prompt. To reflect the higher
operational cost of missed anomalies, \textit{Anomaly} completion tokens receive
increased loss weight. Let $\mathcal{T}$ denote the response-token positions,
where $\omega_t=50$ for anomaly-response tokens and $\omega_t=1$ otherwise. The
training objective is

\begin{equation}
\mathcal{L}_{\mathrm{SLM}}
=
-\frac{1}{|\mathcal{T}|}
\sum_{t\in\mathcal{T}}
\omega_t
\log p_{\theta}(y_t\mid y_{<t}, W_j).
\end{equation}

% ==============================================================================================================

\subsection{Anomaly Scenario Definitions}

The practitioner-informed scenarios comprise controlled HSM monitoring
conditions motivated by cryptographic-hardware attack research, stealth
reconnaissance strategies, and the stateful object-management semantics of
PKCS\#11~\cite{PKCS11,Bardou2012PaddingOracle,Dabbagh2011SlowScan,
Bortolozzo2010PKCS11,Delaune2010PKCS11}. The scenarios also include a
deployment-specific policy condition grounded in PKCS\#11 operational
constraints~\cite{Centenaro2012PKCS11}. They are not confirmed production
incident traces or a fixed taxonomy from prior HSM audit-log studies.

\begin{itemize}

\item \textit{Cryptographic Failure Probing:} Repeated cryptographic requests
produce concentrated related failures for a common target, reflecting
error-conditioned probing behavior observed in attacks against cryptographic
hardware~\cite{Bardou2012PaddingOracle}.

\item \textit{Object Handle Probing:} PKCS\#11 object-search operations return
handles for matching token or session objects~\cite{PKCS11}. Repeated searches,
accesses, or invalid-handle attempts involving workflow-unjustified objects
indicate possible reconnaissance of token-resident assets~\cite{Bortolozzo2010PKCS11}.

\item \textit{Lifecycle Violation:} A client reuses an object handle after a
destructive operation invalidates its associated object. This scenario tests
whether monitoring preserves object-state progression across events~\cite{PKCS11}.

\item \textit{Policy Violation:} A syntactically valid operation conflicts with
a deployment-specific policy. It represents a policy-governed violation rather
than a generic public attack family~\cite{Centenaro2012PKCS11}.

\item \textit{Low-and-Slow Metadata Probing:} A client intermittently queries
unrelated keys or objects during routine activity, adapting low-rate stealth
reconnaissance strategies to HSM metadata access~\cite{Dabbagh2011SlowScan}.

\end{itemize}

Together, these scenarios test whether HSMLog preserves the key/object,
session, state, and result relationships needed to recognize HSM-specific
behavior.

% ==============================================================================================================

\subsection{Baselines}

We compare HSMLog with DeepLog~\cite{Du2017DeepLog},
LogRobust~\cite{Zhang2019LogRobust},
PLELog~\cite{Yang2021PLELog},
LogBERT~\cite{Guo2021LogBERT},
NeuralLog~\cite{Le2021NeuralLog},
LogLLM~\cite{Guan2024LogLLM}, and
MIDLog~\cite{He2025MIDLog}.

These methods span sequential prediction, robust representations, parsing-free
semantic encoding, weak supervision, self-supervision, and language-model-based
analysis. They do not use HSMLog's policy-guided assessment, retrieval review,
or behavioral-analysis components. This comparison evaluates HSMLog against generic methods in the studied HSM
setting.

Policy Violation is an unseen policy-driven test category. Baselines assess it
with native scoring or classification because they do not receive HSMLog's
organization-specific policies.

All baselines were trained and evaluated using the same session-disjoint data
splits and case-aware incident protocol as HSMLog. Their configurations and
decision thresholds were fixed before final-test evaluation.

% ==============================================================================================================

\subsection{Case-Aware Alert-Incident Evaluation}

All methods use a common case-aware alert-incident protocol with one-to-one
incident-case matching, preventing multiple credits for repeated alerts.

Cryptographic Failure Probing, Object Handle Probing, and Policy Violation
require at least 50\% case-event coverage. Lifecycle Violation requires 75\%
coverage for its state-transition pattern. Low-and-Slow Metadata Probing
requires at least five overlapping events, or all events in shorter cases.

Case detection rate (CDR) is the fraction of anomaly cases matched by an alert
incident. Event coverage (EC) is the fraction of anomalous events covered by
matched incidents. Precision, recall, and F1 are computed over alert incidents.

% ==============================================================================================================

\section{Results and Discussion}

\subsection{Comparison with Representative Baselines}

Table~\ref{tab:main_results} reports category-wise CDR and EC together with
overall EC, precision, recall, and F1.

\begin{table*}[t]
\centering
\caption{Case detection and anomalous-event coverage by category, with
overall incident-matching performance.}
\label{tab:main_results}

\footnotesize
\renewcommand{\arraystretch}{1.24}
\setlength{\tabcolsep}{1.5pt}

\begin{threeparttable}
\begin{tabular*}{\textwidth}{
@{\extracolsep{\fill}}
l
*{14}{M}
@{}
}
\toprule

\multirow{2}{*}{\textbf{Method}}
& \multicolumn{2}{c}{\makebox[0pt][c]{\shortstack{\textbf{Cryptographic}\\\textbf{Failure Probing}}}}
& \multicolumn{2}{c}{\makebox[0pt][c]{\shortstack{\textbf{Object Handle}\\\textbf{Probing}}}}
& \multicolumn{2}{c}{\makebox[0pt][c]{\shortstack{\textbf{Lifecycle}\\\textbf{Violation}}}}
& \multicolumn{2}{c}{\makebox[0pt][c]{\shortstack{\textbf{Policy}\\\textbf{Violation}}}}
& \multicolumn{2}{c}{\makebox[0pt][c]{\shortstack{\textbf{Low-and-Slow}\\\textbf{Metadata Probing}}}}
& \multicolumn{4}{c}{\makebox[0pt][c]{\textbf{Overall}}} \\

\cmidrule(lr){2-3}
\cmidrule(lr){4-5}
\cmidrule(lr){6-7}
\cmidrule(lr){8-9}
\cmidrule(lr){10-11}
\cmidrule(l){12-15}

& \textbf{CDR} & \textbf{EC}
& \textbf{CDR} & \textbf{EC}
& \textbf{CDR} & \textbf{EC}
& \textbf{CDR} & \textbf{EC}
& \textbf{CDR} & \textbf{EC}
& \textbf{EC} & \textbf{Prec.} & \textbf{Rec.} & \textbf{F1} \\

\midrule

DeepLog
& 90.48 & 92.15
& 89.47 & 94.89
& 50.00 & 79.46
& 90.48 & \textbf{100.00}
& 89.47 & 96.60
& 93.26 & 93.18 & 82.00 & 87.23 \\

LogRobust
& 80.95 & 83.14
& \textbf{100.00} & \textbf{100.00}
& 65.00 & 93.80
& 90.48 & \textbf{100.00}
& \textbf{94.74} & 96.94
& 94.72 & 90.53 & 86.00 & 88.21 \\

PLELog
& 95.24 & 95.93
& 94.74 & 95.21
& 60.00 & 60.47
& \textbf{100.00} & \textbf{100.00}
& \textbf{94.74} & 94.22
& 90.59 & 96.74 & 89.00 & 92.71 \\

LogBERT
& 95.24 & 95.93
& 94.74 & 95.21
& 60.00 & 60.47
& \textbf{100.00} & 98.35
& \textbf{94.74} & 96.60
& 90.65 & 93.68 & 89.00 & 91.28 \\

NeuralLog
& 95.24 & \textbf{97.97}
& \textbf{100.00} & \textbf{100.00}
& 60.00 & \textbf{100.00}
& \textbf{100.00} & 96.97
& 89.47 & \textbf{100.00}
& \textbf{98.85} & 94.68 & 89.00 & 91.75 \\

LogLLM
& 90.48 & 90.99
& 89.47 & 90.10
& 60.00 & 60.47
& 85.71 & 87.33
& 84.21 & 85.71
& 83.97 & 94.25 & 82.00 & 87.70 \\

MIDLog
& 85.71 & 86.63
& 89.47 & 90.10
& 55.00 & 54.26
& 85.71 & 87.33
& 84.21 & 85.71
& 82.00 & \textbf{100.00} & 80.00 & 88.89 \\

\specialrule{0.8pt}{1.2pt}{1.2pt}

\textbf{HSMLog}
& \textbf{100.00} & 95.06
& 94.74 & \textbf{100.00}
& \textbf{90.00} & 98.45
& \textbf{100.00} & \textbf{100.00}
& \textbf{94.74} & \textbf{100.00}
& 98.66 & 98.97 & \textbf{96.00} & \textbf{97.46} \\

\bottomrule
\end{tabular*}

\begin{tablenotes}[flushleft]
\footnotesize
\item[] \emph{CDR} = case detection rate; \emph{EC} = anomalous-event
coverage. All values are percentages. Boldface indicates the best value in each
column.
\end{tablenotes}

\end{threeparttable}
\end{table*}

HSMLog achieves the highest recall and F1 score: 96.00\% recall, 98.66\% EC,
98.97\% precision, and a 97.46\% F1 score. PLELog is the strongest baseline
in terms of F1 at 92.71\%; HSMLog improves F1 by 4.75 points and recall by
7 points.

EC alone is insufficient: NeuralLog attains 98.85\% EC, 0.19 points above
HSMLog, but only 89.00\% recall and 91.75\% F1. HSMLog detects seven more cases
while retaining 98.66\% EC and improving precision from 94.68\% to 98.97\%.

% ==============================================================================================================

\subsection{Contribution of Contextual Detection, Policy-Guided Assessment,
and Retrieval-Grounded Candidate Review}

Table~\ref{tab:ablation_window}(a) isolates the contributions of contextual
SLM detection, policy-guided assessment, and retrieval-based candidate review.

\begin{table*}[!t]
\centering
\caption{Component ablation and observation-window sensitivity of HSMLog.}
\label{tab:ablation_window}
\vspace{-0.7em}

\noindent
\begin{minipage}[t]{0.59\textwidth}
\vspace{0pt}
\footnotesize
{\centering\textbf{(a) Component ablation}\par}
\vspace{1em}

\renewcommand{\arraystretch}{1.08}
\setlength{\tabcolsep}{1pt}

\begin{tabular*}{\linewidth}{
@{\extracolsep{\fill}}
lcccc
@{}
}
\toprule
\textbf{Configuration} &
\textbf{EC} &
\textbf{Prec.} &
\textbf{Rec.} &
\textbf{F1} \\
\midrule

Contextual SLM only
& 68.70 & 84.78 & 78.00 & 81.25 \\

Policy-guided assessment only
& 23.09 & \textbf{100.00} & 21.00 & 34.71 \\

Stage~1: Contextual + Policy-guided
& \textbf{98.66} & 97.96 &
\textbf{96.00} & 96.97 \\

\textbf{Full HSMLog: Stage~1 + Retrieval Review}
& \textbf{98.66} & 98.97 &
\textbf{96.00} & \textbf{97.46} \\

\bottomrule
\end{tabular*}
\end{minipage}%
\hfill
\begin{minipage}[t]{0.36\textwidth}
\vspace{0pt}
\footnotesize
\centering
\textbf{(b) Stage~1 window sensitivity}\par
\vspace{1em}

\renewcommand{\arraystretch}{1.08}
\setlength{\tabcolsep}{0.3pt}

\begin{tabular*}{\linewidth}{
@{\extracolsep{\fill}}
ccccc
@{}
}
\toprule
\textbf{$w$} &
\textbf{EC} &
\textbf{Prec.} &
\textbf{Rec.} &
\textbf{F1} \\
\midrule

10
& 96.12 & 98.99 &
\textbf{98.00} & \textbf{98.49} \\

\textbf{20}
& \textbf{98.66} & 97.96 &
96.00 & 96.97 \\

25
& 78.63 & \textbf{100.00} &
70.00 & 82.35 \\

30
& 78.18 & \textbf{100.00} &
69.00 & 81.66 \\

50
& 86.07 & \textbf{100.00} &
64.00 & 78.05 \\

\bottomrule
\end{tabular*}
\end{minipage}

\end{table*}

Contextual SLM alone achieves 78.00\% recall, 68.70\% EC, and 81.25\% F1,
showing contextual patterns from structured windows but incomplete
coverage. Policy-guided assessment alone achieves 100.00\% precision but
21.00\% recall, showing policy knowledge cannot cover all evaluated patterns.

Combining both raises recall to 96.00\%, EC to 98.66\%, and F1 to
96.97\%, improving recall by 18 points over contextual SLM alone.

Retrieval review raises precision from 97.96\% to 98.97\% and F1 from 96.97\%
to 97.46\%, with unchanged recall and EC. The 1.01-point precision gain
and 0.49-point F1 gain reflect Stage~2's bounded triage role, not a second detector.

% ==============================================================================================================

\subsection{Observation-Window Trade-off}

Table~\ref{tab:ablation_window}(b) reports a post-hoc sensitivity analysis of
Stage~1 under alternative observation-window sizes.

Although $w=10$ yields higher recall and F1, it provides lower anomalous-event
coverage than $w=20$. We fixed $w=20$ before final-test evaluation to favor
broader preservation of candidate-related evidence, which is important for
subsequent retrieval-grounded review and behavioral analysis, rather than maximizing F1 alone.
The results therefore illustrate a trade-off between incident detection
performance and anomalous-event coverage. Beyond $w=20$, recall drops as
unrelated background dilutes localized evidence.

% ==============================================================================================================

\subsection{Role of Retrieval-Grounded Behavioral Analysis}

For each retained incident, HSMLog organizes retrieved policy evidence,
pre-incident suspicious-key context, and the associated candidate event sequence
into a retrieval-grounded behavioral analysis. The analysis identifies the
affected key or object, summarizes suspicious progression, states supporting
evidence, and highlights investigation-relevant context.

The analysis follows Stage~1 detection and Stage~2 review rather than serving
as a free-form additional detector. We evaluate Stage~2 as a bounded
candidate-review mechanism, not as a substitute for engineer judgment. The
generated analysis supports post-detection explanation and triage.

% ==============================================================================================================

\section{Related Work}

DeepLog~\cite{Du2017DeepLog} models normal sequences;
LogBERT~\cite{Guo2021LogBERT} learns contextual representations through masked
language modeling; LogRobust~\cite{Zhang2019LogRobust} targets unstable or
noisy logs; and NeuralLog~\cite{Le2021NeuralLog} uses neural representations
to reduce explicit-template dependence.

PLELog~\cite{Yang2021PLELog} and MIDLog~\cite{He2025MIDLog} address limited or
inexact supervision, while LogLLM~\cite{Guan2024LogLLM} explores
language-model-based log analysis.

HSMLog applies an SLM directly to structured HSM windows rather than only for
post-hoc explanation, combines learned context with policy-guided assessment,
and uses retrieval-grounded review and analysis to connect detection with
operational investigation.

% ==============================================================================================================

\section{Lessons Learned and Limitations}

Three practitioner lessons emerge. First, HSM monitoring should preserve
key/object, session, state, and timing context. Second, learned contextual
detection benefits from deployment-specific policy knowledge. Third,
post-detection review should remain conservative and evidence-grounded.
These principles may extend to other stateful, policy-governed systems, including
key-management and access-control systems and regulated workflows, with
domain-specific log representations and policies.

This study has limitations. Normal background data derive from real industrial
operation, whereas anomaly cases are practitioner-informed scenarios co-defined
with industrial partners and embedded in the evaluation stream; they are
meaningful HSM anomaly conditions, not confirmed production incidents.
Evaluation is limited to one HSM environment and structured log schema; some
policy conditions are only functionally described to protect confidential
operational knowledge. Stage~2's usefulness to HSM engineers was not directly
evaluated; future work should study this with practitioners and additional HSM
vendors, workloads, and policies.

% ==============================================================================================================

\section{Conclusion}

This paper presented HSMLog, a two-stage framework that combines SLM-based
contextual detection and policy-guided assessment over structured HSM windows
with retrieval-grounded candidate review and behavioral analysis. On real
industrial HSM background logs augmented with anomaly scenarios co-defined
with industrial partners, HSMLog achieves 98.97\% precision, 96.00\% recall,
98.66\% anomalous-event coverage, and a 97.46\% F1 score, demonstrating effective
alerting and incident triage in the studied setting.

% ==============================================================================================================

\section*{Acknowledgment}
This work was conducted with technical guidance and support from eMemory Technology Inc. and PUFsecurity.

% ==============================================================================================================

\bibliographystyle{IEEEtran}
\bibliography{references}

\end{document}